# Data Mining as a Torch Bearer in Education Sector

**Umesh Kumar Pandey[1], Brijesh Kumar Bhardwaj[2], Saurabh pal[3]**

[1,2] **Research scholar, Singhania Univeraity
Pacheri Bari Jhunjhunu Rajasthan**
Umesh6326@gmail.com[1], pandeyu2003@yahoo.co.in[1], wwwbkb@rediffmail.com[2]

[3] **Head, Dept. of Computer Applications,
VBS Purvanchal University
Jaunpur, UP**
drsaurabhpal@yahoo.co.in

## ABSTRACT

Every data has a lot of hidden information. The processing method of data decides what type of information data produce. In India education sector has a lot of data that can produce valuable information. This information can be used to increase the quality of education. But educational institution does not use any knowledge discovery process approach on these data. Information and communication technology puts its leg into the education sector to capture and compile low cost information. Now a day a new research community, educational data mining (EDM), is growing which is intersection of data mining and pedagogy. In this paper we present roadmap of research done in EDM in various segment of education sector.

## INTRODUCTION

Quality education is one of the most promising responsibilities of any country to his countrymen. Quality education does not mean high level of knowledge produced. But it means that education is produced to students in efficient manner so that they learn without any problem. For this purpose quality education includes features like: methodology of teaching, continuous evaluation, categorization of student into similar type, so that students have similar objectives, demographic, educational background etc.

Advent of computer opens a new era in the field of database because of high storage capability and complex study. Huge number of data can be organized in any order on just a click of mouse. It can explore a new knowledge on these data which was either impossible for a human mind or a very time consuming process.

Education sector has a lot of data in the form student information. Application of computer in the education can extract valuable information to provide quality education. Due to this combination, of education and computer (data mining) a new research community is growing i.e. educational data mining [17].

## Motivational work

Educational data mining is a technological step in the education sector. It provides a new way of look into the education which was hidden from humankind. C romero ans S vetura [17] made a comprehensive study on the development of this educational data mining since 1995 to 2005. Their paper surveys the application of data mining to traditional education systems, particular web-based course, well known learning content management systems and adaptive and intelligent web-based educational systems.

Concluding from different research paper they wrote that the application of the application of data mining in educational system have different objective at student, educators and academics responsible and administrators. From students orientation its objective is to recommend to learners activities, resources, learning tasks, suggest path pruning etc. From educators point of view its objective is to get more objective feedback, effectiveness on learning process, monitoring, find learners mistake etc. an academic responsible and administrator's objective to use it to improve site, efficiency, better organize institutional resources, enhance educational program etc.

| Authors | Mining task | Educational system |
|---|---|---|
| Sanjeev and Zytkow (1995) | Sequence pattern | Traditional education |
| Zaïane et al. (1998) | Statistic and sequence pattern | LCM systems |
| Beck and Woolf (2000) | Prediction | AIWBE system |
| Becker et al. (2000) | Association and classification | Traditional education |
| Chen et al. (2000) | Classification | Web-based course |
| Ha et al. (2000) | Association | Web-based course |
| Ma et al. (2000) | Association | Traditional education |
| Tang et al. (2000) | Text mining | AIWBE system |
| Yu et al. (2001) | Association | Web-based course |
| Zaïane and Luo (2001) | Sequence pattern | LCM system |
| Luan (2002) | Clustering and prediction | Traditional education |
| Pahl and Donnellan (2003) | Sequence pattern and statistics | LCM system |
| Shen et al. (2002) | Visualization | LCM system |
| Wang (2002) | Association and sequence pattern | Web-based course |
| Merceron and Yacef (2003) | Statistic | AIWBE system |
| Minaei-Bidgoli and Punch (2003) | Classification | Web-based course |
| Shen et al. (2003) | Sequence pattern and clustering | Web-based course |
| Zarzo (2003) | Statistic | Web-based course |
| Arroyo et al. (2004) | Prediction | AIWBE system |
| Baker et al. (2004) | Classification | AIWBE system |
| Chen et al. (2004) | Text mining | Web-based course |
| Freyberger et al. (2004) | Association | AIWBE system |
| Hamalainen et al. (2004) | Classification | AIWBE system |
| Heiner et al. (2004) | Statistic | AIWBE system |
| Lu (2004) | Association | AIWBE system |
| Merceron and Yacef (2004) | Association | AIWBE system |
| Minaei-Bidgoli et al. (2004) | Association | Web-based course |
| Mor and Minguillon (2004) | Clustering | LCM system |
| Romero et al. (2004) | Association | AIWBE system |
| Talavera and Gaudioso (2004) | Clustering | LCM system |
| Ueno (2004b) | Outlier detection | Web-based course |
| Ueno (2004a) | Text mining | Web-based course |
| Wang et al. (2004) | Sequence pattern and clustering | LCM system |
| Li and Zaïane (2004) | Association | LCM system |
| Avouris et al. (2005) | Statistic | Web-based course |
| Castro et al. (2005) | Outlier detection | LCM system |
| Dringus and Ellis (2005) | Text mining | LCM system |
| Feng et al. (2005) | Prediction | AIWBE system |
| Hammouda and Kamel (2005) | Text mining | Web-based course |
| Markellou et al. (2005) | Association | Web-based course |
| Mazza and Milani (2005) | Visualization | LCM system |
| Mostow et al. (2005) | Visualization | AIWBE system |
| Muehlenbrock (2005) | Outlier detection | AIWBE system |
| Nilakant and Mitrovic (2005) | Statistic | AIWBE system |
| Tang and McCalla (2005) | Clustering | AIWBE system |
| Zorrilla et al. (2005) | Statistic | LCM system |
| Damez et al. (2005) | Classification | AIWBE system |
| Bari and Benzater (2005) | Text mining | LCM system |

**Table 1:** Works about applying data mining techniques in educational systems [17]

They provide a table (table 1) which contains specific application of data mining technique used by the researchers in their paper. In this paper they provide another table (table 2) which contain some general data mining tools that provide mining algorithms, filtering and visualization techniques

| Tool name | Authors | Mining task |
|---|---|---|
| Mining tool | Zaïane and Luo (2001) | Association and patterns |
| MultiStar | Silva and Vieira (2002) | Association and classification |
| Data Analysis Center | Shen et al. (2002) | Association and classification |
| EPRules | Romero et al. (2003) | Association |
| KAON | Tane et al. (2004) | Text mining and clustering |
| TADA-ED | Merceron and Yacef (2005) | Classification and association |
| O3R | Becker et al. (2005) | Sequential patterns |
| Synergo/ColAT | Avouris et al. (2005) | Statistics and visualization |
| GISMO/CourseVis | Mazza and Milani (2005) | Visualization |
| Listen tool | Mostow et al. (2005) | Visualization |
| TAFPA | Damez et al. (2005) | Classification |
| iPDF_Analyzer | Bari and Benzater (2005) | Text mining |

**Table 2:** Some specific educational data mining, statistics and visualization tools. [17]

Their survey work motivated us to make a study on some research which used data mining technique to find hidden information from educational database. Our focus is on the brief summary of the research, conclusion of the paper and the name of the data mining methodology used in that paper to extract the knowledge.

## Data mining techniques

Hand et al. [9] defined data mining is the analysis of observational data sets to find unsuspected relationships and to summarize the data in novel ways that are both understandable and useful in the data owner.

Dunham [4] said, "Data mining id the use of algorithm to extract the information and patterns derived by the KDD process". He prepares a timeline chart of data mining development.

| Time | Area | Contribution |
|---|---|---|
| Late 1700 | Stat | Bayes theorem of probability |
| Early 1900 | Stat | Regression analysis |
| Early 1920 | Stat | Maximum likelihood estimate |
| Early 1940 | AI | Neural network |
| Early 1950 | | Nearest neighbor |
| Early 1950 | | Single link |
| Late 1950 | AI | Perceptron |
| Late 1950 | Stat | Resampling, bias reduction, Jacckkbnife estimator |
| Early 1960 | AI | ML started |
| Early 1960 | DB | Batch reports |
| Mid 1960 | | Decision tree |
| Mid 1960 | Stat | Linear models of classification |
| | IR | Similarity measure |
| | IR | clustering |
| | Stat | Exploratory data analysis |
| Late 1960 | DB | Relational data model |
| Early 1970 | IR | SMART IR system |
| Mid 1970 | AI | Genetic algorithm |
| Late 1970 | Stat | EM algorithm |
| Late 1970 | Stat | K-means algorithm |
| Early 1980 | AI | Kohonen self-organizing map |
| Mid 1980 | AI | Decision tree algorithm |
| Early 1990 | DB | Association rule |
| 1990 | DB | Data warehousing |
| 1990 | DB | OLAP |

**Table 3**: time line of data mining [4]

We are presenting data mining techniques in following points: classification, clustering and association.

**Classification:** Estimation and prediction may be viewed as types of classification. The problem usually is evaluating the training data set and second apply the model developed [4]. Different classification algorithms are categorized in following table:

| Type | Name of algorithm |
|---|---|
| Statistical | Regression<br>Bayesian |
| Distance | Simple distance<br>K nearest neighbors |
| Decision tree | ID3<br>C4.5<br>CART<br>SPRINT |
| Neural network | Propagation<br>NN supervised learning<br>Radial base function network |
| Rule based | Genetic rules from DT<br>Genetic rules from NN<br>Genetic rules without DT and NN |

**Table 4**: classification algorithm

**Clustering:** Clustering method is grouping of data, which is not predefined. By using clustering technique we can identify dense and sparse regions in object space. Following table provide different clustering technique:

| Type | Name of algorithm |
|---|---|
| Similarity and distance measure | Similarity and distance measure |
| Outlier | Outlier |
| Hierarchical | Agglomerative, Divisive |
| Partitional | Minimum spanning tree<br><br>Squared matrix<br><br>k-means<br><br>nearest neighbor<br><br>PAM<br><br>Bond energy<br><br>Clustering with neural network |
| Clustering large database | BIRCH<br><br>DB Scan<br><br>CURE |
| Categorical | ROCK |

**Table 5**: clustering algorithm

**Association:** The central task of association rule mining [20] is to find sets of binary variables that co-occur together frequently in a transaction database, while the goal of feature selection problem is to identify groups of that are strongly correlated with each other of with a specific target variable.

Association rule has algorithm like: Apriori, CDA, DDA, interestingness measure etc. Tan et al. specified several interestingness measures which is shown in table

| # | Measure | Definition |
|---|---|---|
| 1 | $\phi$-coefficient | $\frac{P(A,B)-P(A)P(B)}{\sqrt{P(A)P(B)(1-P(A))(1-P(B))}}$ |
| 2 | Goodman-Kruskal's ($\lambda$) | $\frac{\sum_j \max_k P(A_j,B_k)+\sum_k \max_j P(A_j,B_k)-\max_j P(A_j)-\max_k P(B_k)}{2-\max_j P(A_j)-\max_k P(B_k)}$ |
| 3 | Odds ratio ($\alpha$) | $\frac{P(A,B)P(\overline{A},\overline{B})}{P(A,\overline{B})P(\overline{A},B)}$ |
| 4 | Yule's $Q$ | $\frac{P(A,B)P(\overline{AB})-P(A,\overline{B})P(\overline{A},B)}{P(A,B)P(\overline{AB})+P(A,\overline{B})P(\overline{A},B)} = \frac{\alpha-1}{\alpha+1}$ |
| 5 | Yule's $Y$ | $\frac{\sqrt{P(A,B)P(\overline{AB})}-\sqrt{P(A,\overline{B})P(\overline{A},B)}}{\sqrt{P(A,B)P(\overline{AB})}+\sqrt{P(A,\overline{B})P(\overline{A},B)}} = \frac{\sqrt{\alpha}-1}{\sqrt{\alpha}+1}$ |
| 6 | Kappa ($\kappa$) | $\frac{P(A,B)+P(\overline{A},\overline{B})-P(A)P(B)-P(\overline{A})P(\overline{B})}{1-P(A)P(B)-P(\overline{A})P(\overline{B})}$ |
| 7 | Mutual Information ($M$) | $\frac{\sum_i \sum_j P(A_i,B_j) \log \frac{P(A_i,B_j)}{P(A_i)P(B_j)}}{\min(-\sum_i P(A_i) \log P(A_i), -\sum_j P(B_j) \log P(B_j))}$ |
| 8 | J-Measure ($J$) | $\max \left( P(A,B) \log(\frac{P(B|A)}{P(B)}) + P(A\overline{B}) \log(\frac{P(\overline{B}|A)}{P(\overline{B})}), \right.$<br>$\left. P(A,B) \log(\frac{P(A|B)}{P(A)}) + P(\overline{A}B) \log(\frac{P(\overline{A}|B)}{P(\overline{A})}) \right)$ |
| 9 | Gini index ($G$) | $\max \left( P(A)[P(B|A)^2 + P(\overline{B}|A)^2] + P(\overline{A})[P(B|\overline{A})^2 + P(\overline{B}|\overline{A})^2] \right.$<br>$-P(B)^2 - P(\overline{B})^2,$<br>$P(B)[P(A|B)^2 + P(\overline{A}|B)^2] + P(\overline{B})[P(A|\overline{B})^2 + P(\overline{A}|\overline{B})^2]$<br>$\left. -P(A)^2 - P(\overline{A})^2 \right)$ |
| 10 | Support ($s$) | $P(A,B)$ |
| 11 | Confidence ($c$) | $\max(P(B|A), P(A|B))$ |
| 12 | Laplace ($L$) | $\max \left( \frac{NP(A,B)+1}{NP(A)+2}, \frac{NP(A,B)+1}{NP(B)+2} \right)$ |
| 13 | Conviction ($V$) | $\max \left( \frac{P(A)P(\overline{B})}{P(A\overline{B})}, \frac{P(B)P(\overline{A})}{P(B\overline{A})} \right)$ |
| 14 | Interest ($I$) | $\frac{P(A,B)}{P(A)P(B)}$ |
| 15 | cosine ($IS$) | $\frac{P(A,B)}{\sqrt{P(A)P(B)}}$ |
| 16 | Piatetsky-Shapiro's ($PS$) | $P(A,B) - P(A)P(B)$ |
| 17 | Certainty factor ($F$) | $\max \left( \frac{P(B|A)-P(B)}{1-P(B)}, \frac{P(A|B)-P(A)}{1-P(A)} \right)$ |
| 18 | Added Value ($AV$) | $\max(P(B|A) - P(B), P(A|B) - P(A))$ |
| 19 | Collective strength ($S$) | $\frac{P(A,B)+P(\overline{AB})}{P(A)P(B)+P(\overline{A})P(\overline{B})} \times \frac{1-P(A)P(B)-P(\overline{A})P(\overline{B})}{1-P(A,B)-P(\overline{AB})}$ |
| 20 | Jaccard ($\zeta$) | $\frac{P(A,B)}{P(A)+P(B)-P(A,B)}$ |
| 21 | Klosgen ($K$) | $\sqrt{P(A,B)} \max(P(B|A) - P(B), P(A|B) - P(A))$ |

**Figure 1**: interestingness measure of association[20]

**Fact analysis in research papers**

Elena susnea [19] concluded that development of come adequate and efficient teaching strategy is not a simple operation. It implies a contextual, original and unique combination of the entire training process. Elena used K-means clustering data mining technique on 5 institutions of higher

119

education and realized 3 clusters taking into account the study year and the answer to question "how much do the teaching objectives stated in the educational plan and analytical program correspond to your operation".

Oladipupo and oyelade [12] made their study using association rule data mining technique to identify student's failure patterns. They take a total number of 30 courses for 100 level and 200 level. Their study focuses on constructive recommendation, curriculum structure and modification in order to improve student's academic performance and trim down failure rate.

Erdogan and Timor [7] conducted their study using five fields i.e. area point percent, success grade, sexcode, high school type ID and faculty ID with responses yes/no. they find the relationship between students university entrance examination results and their success was studied using K-means clustering data mining method.

Merceron and Yacef [10] made a case study for providing additional resources to students in a face to face teaching context. Teacher wants to figure out whether these resources and possibly whether their use has any (positive) impact on marks. They concluded association rules are useful in educational data mining. This technique requires not only support and confidence but also measures of interestingness be considered to retain meaningful rules and filter interestingness ones out. Interestingness should be checked with cosine first and then with lift. If both measure disagree, teacher should use the intuition behind the measure to decide whether or not to dismiss the association rule.

Pandey and Pal [13] conducted a study to analyze the impact of language on the presence of students in class room. For this paper association rule is use which finds the appropriate language and attendance in the classroom. Data is collected from PSRIET, pratapgarh UP India, PGDCA of computer science department. Data size was 60 students. On measuring several interestingness measure they concluded that the students are looking for mixed (English and Hindi) mode class rather that pure Hindi an English.

Suthan and Baboo [18] proposed CHAID algorithm for framework multi dimensional student assessment (MUSTAS) to measure student's performance through multidimensional attribute. MUSTAS framework consist of demographic factors, academic performance and dimensional factor, further subdivided into three dimensions respectively self assessment, institutional assessment and external assessment. These dimensional factors help to measure the student's attitude.

Falakmasir and Habibi's [8] research is aimed to investigate the impact of a number of e-learning activities on the student's learning development. For this purpose feature ranking and attribute selection methods have been propose and different metrics to discard irrelevant feature and select the important ones including: information gain, gain ration, symmetrical uncertainty, relief-F, one, R and chi squared used non virtual class room, archive view, forum read, assignment view, assignment load discussion read, resource view, forum post, discussion post. The main idea is to rank the learning activities based on their importance in order to improve student's performance on the most important. A decision tree was created using C4.5 algorithm by result obtained from the research. They concluded that participation in virtual classroom sessions has the greatest impact on effectiveness of learning in their particular settings of IUST e-learning.

Bhardwaj and Pal [3] collected data of 50 students from VBSP University, UP India from MCA department from session 2007 to 2010. They gathered student related data in following variable: previous semester marks, class test grade, seminar performance, assignment, general proficiency, attendance, Lab work, end semester marks. On the available information in these variables they apply ID3 algorithm to draw the decision tree. This will help to the students and the teachers to improve the division of the student.

Ayesha et al. [2] used K-means clustering data mining technique on 120 students of department of Computer Science University of agriculture Faisalabad in 2008-09 to analyze students learning behavior. The student's evaluation factor like class quizzes, mid and final exam assignment are studied. The proposed model identifies the weak students before final in order to save them from serious harm. Teacher's can use this analysis to improve the performance of students.

Ramasubramaniam et al. [15] analyze student information system (SIS) using rough set theory to predict the future of students. The rough set theory is a recent mathematical theory employed as a data mining tool with many favorable advantages. They analyzed student information system with the attributes such as academic, non-academic and human behavior relationship. Anyone can use this technique to access the performance of a particular student by accessing a group of students even if they do not aware of certain attribute values of the students.

Pandey and Pal [14] analyzed 600 students data collected from PGDCA course of RMLA

University UP India. They transformed their data on three criteria caste category (Gen, OBC, SC/ST), language medium (Hindi, English) and class ( BA(NC), BA(CA), BSc(Bio), BSc( math) and BCom), each is divided into first, second, third and fail category. Naïve Bayes classification and prediction methodology is used on these data. In this study they predicted that, what is success ratio of particular class, medium and caste student.

Merceron and Yacef [11] were interested in detecting association mistakes done by student using interestingness measures such as lift, conviction, correlation etc. They explored the interestingness measures under different variant of the data sets. In this paper they took 230 students who are attempting 2000 exercises and making mistakes X and Y in four possible order ( (y, y);(y, n);(n, y);(n, n)). They first extracted association rule from 2002 and 2001 data. In 2003, after the end of semester, mining for mistakes association was conducted again. They concluded that result did not much changed.

Ramaswami and Bhaskaran [16] focused on different factors which affecting student's academic performance at higher secondary level. They used CHAID classification tree as a technique to design the student performance prediction model. They concluded in their research paper that features like medium of instruction, marks obtained in secondary education were the strongest indicator for the student performance in higher secondary education.

Zhang and Clark [21] used three algorithms: Naïve Bayes, support vector machine and decision tree to predict student dropout. In order to increase student retention they understand why students dropout. Their student data include average mark, online learning system information, library information, nationality, university entry certificate, course award, current study level, study mode, re_sit number, current year, age, gender, race etc. In this paper they discussed how data mining can help spot students 'at risk', evaluate the course or module suitability and tailor the intervention to increase student retention.

Ajayi et al. [1] worked their research on 260 students from 10 public secondary schools of Ogun sate, Nigeria, to investigate the predictive validity of mathematics mock result of students in SSCE. Statistical software package for social sciences (SPSS), analysis tool is used and findings used to prdict the success in academic performance of students in SSCE Mathematics. They concluded that students performed better in mock examination than in SSCE exam. Their analysis also revealed that there is no gender in performance of boys and girls in both mock and SSCE mathematic.

Abd-eirnamman et al [6] introduced the teaching evaluation (TEI) as an index to quantify students textual evaluation using the number of positive and negative comments interpreted from the text. The answer evaluations were scored positively or negatively in two independent ways: manually using human interpretation, on five major elements of teaching: course, instructor, assignment, material and delivery, and automatically, using wordstat software based on keyword co-occurrence text mining algorithm. Their result proved that text mining is a promising technique to analyze short answer textual information in the students course evaluation sheets more efficiently than by simply having read each comment individually examining the TEI computed from manual interpretation of student results showed significant correlation with student answers to the overall course and instructors evaluation questions located at the front page of the sheet.

ElAlfi et al. [5] explored how data mining is being used in education services at Taif University in Saudi Arabia. In their research process the determined hypothesizes and after testing it with several statistical measure they concluded: superiority of females in logical reasoning ability to understand the linkage of precondition and conclusion; females are more likely to use the correct methods of learning, more able to manage time, take of observation notes and summaries; females have external incentives which lead them to exert effort such as the motives of the desire to challenge the male society significantly, as if to prove a kind of self motivation, ambition and self-reliance; males interact better than female in t he classroom, particularly in t he dimension's of teachers positivity and potential of the classroom. At the end of the proposed algorithm for extracting a set of accurate and comprehensible rules from the input database via trained ANN using genetic algorithm (GA). Study of ElAlfi et al. combines the variable related to personality, mental and environmental aspects in order to reach an integrated view of the learning nature process and the factors affecting it.

## Conclusion and future work

Data mining extracts hidden information with the help different mining technique. Prediction, results and recommendation are provided by this information, which help the user to take further decision. It also guides the concern person for whom information has been extracted. In this paper we discussed different researches did in the education sector using data mining.

Students, educators and academic responsible person can use these findings to improve the quality of education.

## AUTHORS PROFILE

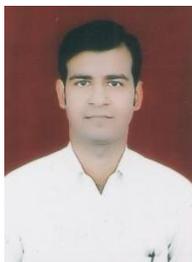
Umesh Kumar Pandey is head and Assistant Professor in the Department of Computer Applications, P S R I E T, Pratapgarh, UP India. He obtained his M.C.A degree from IGNOU (2004) and M.Phil. in Computer Science from PRIST University, Tamilnadu. He is currently doing research in Data Mining and Knowledge Discovery from Singhaniya University, Rajasthan. Umesh Kumar Pandey published two international Journals papers (in IJCSI and IJCSIT). He is member of two international associations i.e. IAENG and IACSIT.

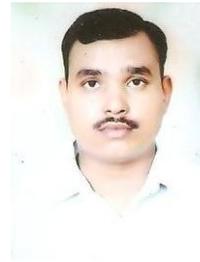
Brijesh Kumar Bhardwaj is Assistant Professor in the Department of Computer Applications, Dr. R. M. L. Avadh University Faizabad India. He obtained his M.C.A degree from Dr. R. M. L. Avadh University Faizabad (2003) and M.Phil. in Computer Applications from Vinayaka mission University, Tamilnadu. He is currently doing research in Data Mining and Knowledge Discovery. He has published one international paper. (IJACSA) International Journal of Advanced Computer Science and Applications,

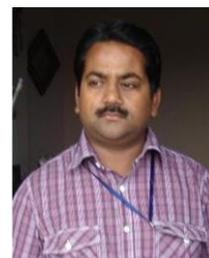
Saurabh Pal received his M.Sc. (Computer Science) from Allahabad University, UP, India (1996) and obtained his Ph.D. degree from the Dr. R. M. L. Awadh University, Faizabad (2002). He then joined the Dept. of Computer Applications, VBS Purvanchal University, Jaunpur as Lecturer. At present, he is working as Head and Sr. Lecturer at Department of Computer Applications.

Saurabh Pal has authored a commendable number of research papers in international/national Conference/journals and also guides research scholars in Computer Science/Applications. He is an active member of CSI, Society of Statistics and Computer Applications and working as reviewer for more than 10 international journals. His research interests include Image Processing, Data Mining, Grid Computing and Artificial Intelligence.